\documentclass{aa}  

\usepackage[breaklinks=true,colorlinks,citecolor=blue]{hyperref}
\usepackage{graphicx}
\usepackage{enumerate}
\usepackage{epsfig}
\usepackage{color}
\usepackage{txfonts}
\usepackage{natbib}
\usepackage{multirow}
\usepackage{ulem} 

\definecolor{ao}{rgb}{0.0, 0.5, 0.0}

\newcommand{\FeH}{\ensuremath{\rm [Fe/H]}}
\newcommand{\Rgini}{\ensuremath{R^{\rm ini}_{\rm g}}\,}
\newcommand{\Rgend}{\ensuremath{R^{\rm end}_{\rm g}}\,}

\newcommand{\kmps}{km~s\ensuremath{^{-1} }\,}

\newcommand{\Msun}{M\ensuremath{_\odot}\,}

\begin{document} 

\title{Escapees from the bar resonances} 
\subtitle{Presence of low-eccentricity metal-rich stars at the solar vicinity}
\titlerunning{Migration of inner disk stars with the bar resonances}

\author{S.~Khoperskov$^{1}$, P.~Di Matteo$^{2,3}$, M.~Haywood$^{2,3}$, A.~G\'{o}mez$^{2}$, O. N. Snaith$^{2}$}
\authorrunning{Khoperskov et al.}

\institute{$^{1}$ Max-Planck-Institut f\"{u}r extraterrestrische Physik, Gie{\ss}enbachstrasse 1, 85748 Garching, Germany  \\$^{2}$ GEPI, Observatoire de Paris, PSL Universit{\'e}, CNRS,  5 Place Jules Janssen, 92190 Meudon, France \\ $^{3}$ Sorbonne Universit\'{e}, CNRS UMR 7095, Institut d'Astrophysique de Paris, 98bis bd Arago, 75014 Paris, France}

 
\abstract{Understanding radial migration is a crucial point for building relevant chemical and dynamical evolution models of the Milky Way disk. In this paper we analyze a high-resolution $N$-body simulation of a Milky Way-type galaxy to study the role that the slowing down of a stellar bar has in generating migration from the inner to the outer disk. Stellar particles are trapped by the main resonances~(corotation and outer Lindblad resonance, OLR) which then propagate  outward across the disk due to the bar slowing down. Once the bar strength reaches its maximum amplitude, some of the stars delivered to the outer disk escape the resonances and some of them settle on nearly circular orbits. The number of  escaped stars gradually increases, also due to the decrease in the bar strength when  the boxy/peanut bulge forms. We show that this mechanism is not limited  to stars on nearly circular orbits;  stars initially on more eccentric orbits can also be transferred outward~(out to the OLR location) and can end up on nearly circular orbits. Therefore, the propagation of the bar resonances outward can induce the circularization of the orbits of some of the migrating stars. The mechanism investigated in this paper can explain the presence of metal-rich stars at the solar vicinity and more generally in the outer Galactic disk. Our dynamical model predicts that up to 3\% of stars  between  corotation and the OLR can be formed in the innermost region of the Milky Way. The epoch of the Milky Way bar formation can be potentially constrained by analyzing the age distribution of the most metal-rich stars at the solar vicinity.}

                
\maketitle

\section{Introduction}\label{sec::intro}

It is widely accepted that stars, once formed, do not necessarily stay at their initial birth radius and can migrate across disks~\citep{1965MNRAS.130..125G,1972MNRAS.157....1L}, a phenomenon usually referred to as radial migration. In recent  decades, this phenomenon has had much attention in theoretical and numerical studies of galaxy dynamics. \cite{2002MNRAS.336..785S} proposed that the fast appearance and dissolution of spiral arms is an efficient mechanism for radial migration that was later confirmed in some $N$-body models of transient spiral arms dynamics~\citep[see, e.g.,][]{2008ApJ...684L..79R,2012MNRAS.426.2089R,2016ApJ...818L...6L}. The importance of evolving spiral arm modes is reported by~\cite{2006MNRAS.368..623M},  \cite{2011A&A...527A.147M}, \cite{2011ApJ...730..109F}, \cite{2012MNRAS.422.1363S}, \cite{2014MNRAS.439..623G}, and \cite{2015MNRAS.447.3576D}. The resonance overlap of spiral arms and bar has also been proposed as an efficient driver of radial migration~\citep[see][]{2010ApJ...722..112M,2019ApJ...882..111D}. Gravitational interactions with satellite galaxies can also cause a substantial radial migration, or  stellar diffusion, especially in the outer disk~\citep{2009MNRAS.397.1599Q}. 

Another efficient source of radial motion of gas and stars is   the galactic bar~
\citep{2011A&A...534A..75B,2013MNRAS.436.1479K,2013A&A...553A.102D,2020ApJ...889...81W}.  \cite{2013A&A...553A.102D}, for example, suggested that migration initiated by the bar can cause significant azimuthal abundance variations across the disk. Observations suggest that the effect of the stellar bar is important for generating mixing processes in disk galaxies~\citep{2001AJ....122.1298G, 2011MNRAS.415..709S,2016MNRAS.460.3784S}.  Since bars are the distinctive feature of the majority of local disk galaxies~\citep[e.g.,][]{2000AJ....119..536E, 2000ApJ...529...93K, 2008ApJ...675.1194B,2011MNRAS.411.2026M} including the Milky Way~\citep{okuda77, maihara78, weiland94, dwek95, binney97, babusiaux05, lopez05, rattenbury07, cao13} and are also found at higher redshifts~\citep[e.g.,][]{1996ApJS..107....1A, 2004ApJ...615L.105J, 2014MNRAS.445.3466S}, quantifying their impact on galaxy disks is fundamental for understanding radial migration, in particular in our own Galaxy.  

\begin{figure*}[t!]
\begin{center}
\includegraphics[width=1\hsize]{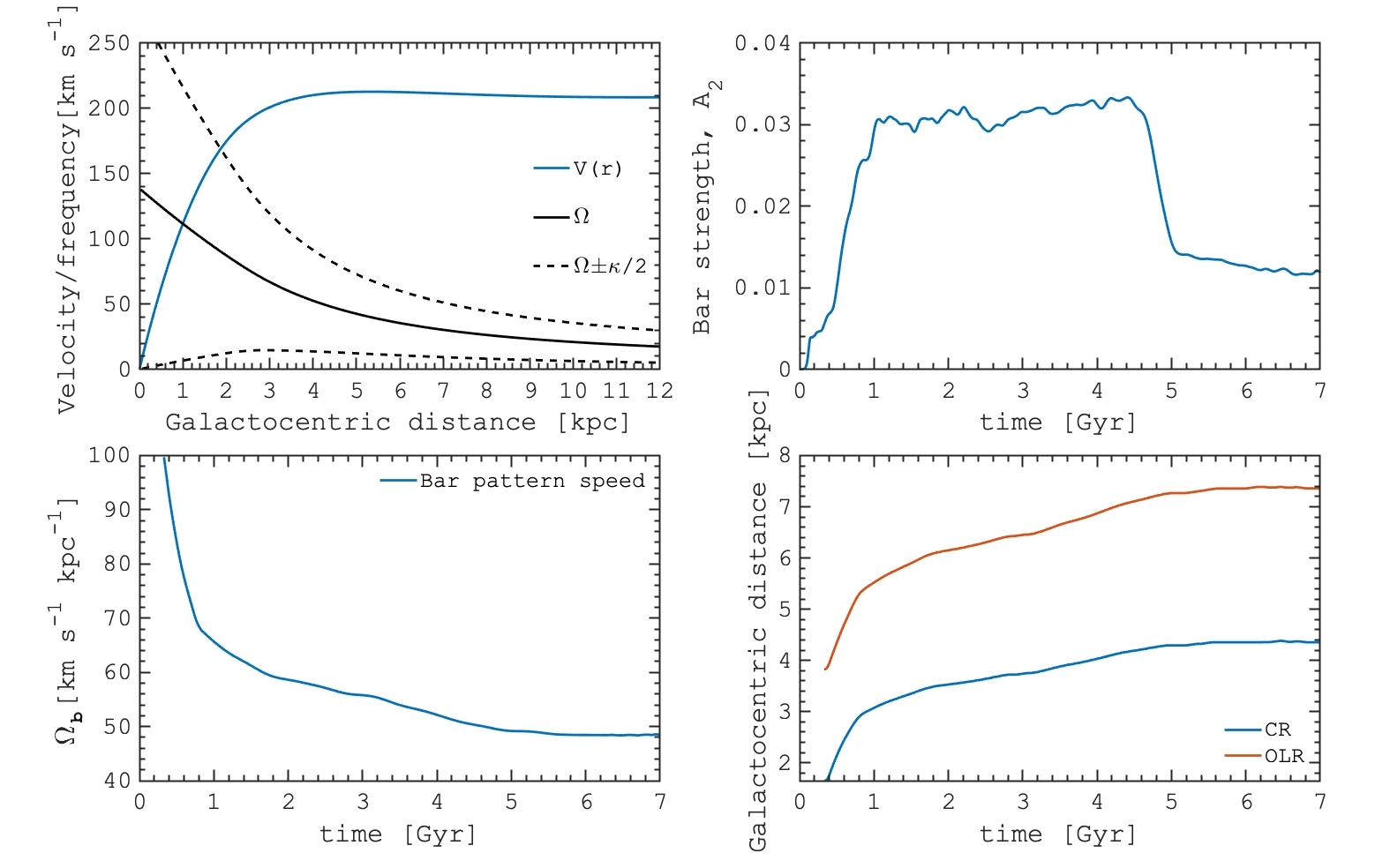}
\caption{Evolution of the main parameters in $N$-body simulation showing the secular evolution of the bar. Rotation curve and main frequencies adopted in the Milky Way-type galaxy model~(top left). The bar in the simulation established at about $\approx1$~Gyr, and due to angular momentum exchange with the dark matter halo later slows down~(bottom left), becoming rounder which reflects in the decrease of the bar strength~($A_2$, top right). Since the bar rotates more slowly the mean location of corotation and the OLR move outward with time~(bottom right). }\label{fig::app1}
\end{center}
\end{figure*}

Pure $N$-body models and collisionless simulations establish that dynamical instabilities and secular changes are accompanied by an increase in  the bar size~\citep{1993A&A...268...65F,2000ApJ...543..704D,2003MNRAS.341.1179A,2010ApJ...719.1470V,2014MNRAS.438L..81A}. Typically in numerical simulations, bars initially grow very rapidly and after possibly passing through a phase of vertical buckling instability, they evolve slowly in a quasi-steady regime~\citep[e.g.,][]{2006ApJ...637..214M}. Bar secular evolution is usually accompanied by an exchange of angular momentum between the different components of a galaxy~(dark halo, bulge, disk). As a consequence, the bar pattern speed tends to decreases with time. For instance, in isolated galaxies the slowing down of the bar is induced by continuous angular momentum transfer from the disk to the surrounding dark matter halo~\citep{2003MNRAS.341.1179A,2003MNRAS.345..406V} and by dynamical friction~\citep{1980A&A....89..296S,2000ApJ...543..704D,2006ApJ...637..567S,2006ApJ...639..868S}. In all cases, the slowing down of the bar inevitably moves the main bar resonances (e.g., inner Lindblad resonance, corotation, and outer Lindblad resonance) toward the outer disk with time~\citep{2018A&A...616A..86H}, and makes the bar longer. 

Among all the stars, those that are affected the most by an asymmetric stellar pattern are found at corotation  (CR), that is  those that co-rotate with the pattern. A steady non-axisymmetric pattern does not result in angular momentum exchange at the CR because stars are trapped on horseshoe orbits~\citep{2002MNRAS.336..785S, 2007MNRAS.379.1155C}. If the pattern is transient or changes significantly  with time, stars can traverse from one side of the CR to another. Hence, for radial migration it is essential that resonances change their location in the disk over time~\citep{2007MNRAS.379.1155C, 2015A&A...578A..58H, 2018A&A...616A..86H}. These last two papers show  that crucially, when the bar is the dominating asymmetry, stars initially within the OLR cannot migrate beyond this resonance.

In this work we report on the results of a numerical study that shows how stars can migrate from the inner to the outer disk of a barred galaxy by following the main bar resonances. 
Building on a foundation by \citet{2015A&A...578A..58H, 2018A&A...616A..86H},  in particular we discuss stars that, trapped at the resonances of a slowing-down stellar bar and hence transported to the outer disk, are found to move on nearly circular orbits once strong migration is over. As detailed in the next sections, the mechanism we discuss in this paper is  able to explain the presence of high-metallicity stars  (i.e., [Fe/H]$> 0.2-0.3$~dex) that are found at the solar vicinity on very low eccentric orbits. These stars have abundances that are characteristic of the inner disk \citep[i.e., $R < 6$~kpc, see][]{2015ApJ...808..132H,2019A&A...625A.105H}, but cold kinematics and  their presence at the solar radius cannot be explained only by blurring~(radial excursion due to epicyclic motions). We suggest that the presence of these high-\FeH stars at the  solar vicinity can be explained in a scenario where the Milky Way experienced the most intense epoch of radial migration at the time of the bar formation. The subsequent decrease in the bar strength, possibly associated with the formation of the boxy/peanut bulge, determined a decrease in the  strength of the migration process. 
The structure of the paper is as follows: in Sect.~\ref{sec::obs} we present the observational motivation for our work based on the analysis of the Gaia-APOGEE sample; in Sect.~\ref{sec::models} a description of the initial conditions and numerical method adopted for the simulation are given; in Sect.~\ref{sec::results} the results including stellar orbits and frequencies analysis presenting the effect of radial migration of the inner disk stars; in Sect.~\ref{sec::discussion} the discussion and main conclusions of this work are presented.

\section{Observational motivation: high-\FeH stars in the solar vicinity}\label{sec::obs}
The presence of old high-metallicity stars~($\FeH> 0.2-0.3$~dex) in the solar vicinity has been investigated for nearly half a century~\citep[see ][]{1972ade..coll...55G}. \cite{1972ade..coll...55G} was one of the first to suggest that these stars probably come from the inner Galaxy. With the chemical cartography provided by recent spectroscopic surveys (APOGEE, LAMOST), it has become  clear that this metal-rich population is dominant not only in the very inner disk of the Milky Way but also  up to a radius of about  R$\sim$6 kpc,  peaking at a metallicity of between $0.2$ and $0.3$~dex~\citep{2015ApJ...808..132H,2019A&A...625A.105H}.  At the solar vicinity, these stars represent only small percentage of the local stellar density, but because these objects have metallicities that exceed those of local star-forming regions, mechanisms that can explain how they  reached the solar orbit and beyond have been investigated \citep[see][]{2002MNRAS.336..785S,2010ApJ...722..112M, 2011A&A...527A.147M}. Blurring (i.e., the increase in the radial oscillation of stars through the increase in their  kinematical orbital energy)   may explain the presence of some of these objects,  but others are on circular orbits with guiding radii of $R\approx8$~kpc or larger, and blurring cannot be invoked in these cases~\citep[][]{2015MNRAS.447.3526K,2018A&A...609A..79H}. Hence, there ought to be a dynamical process that can explain how these objects migrate beyond the inner disk regions. 
\begin{figure}[t!]
\begin{center}
\includegraphics[width=1\hsize]{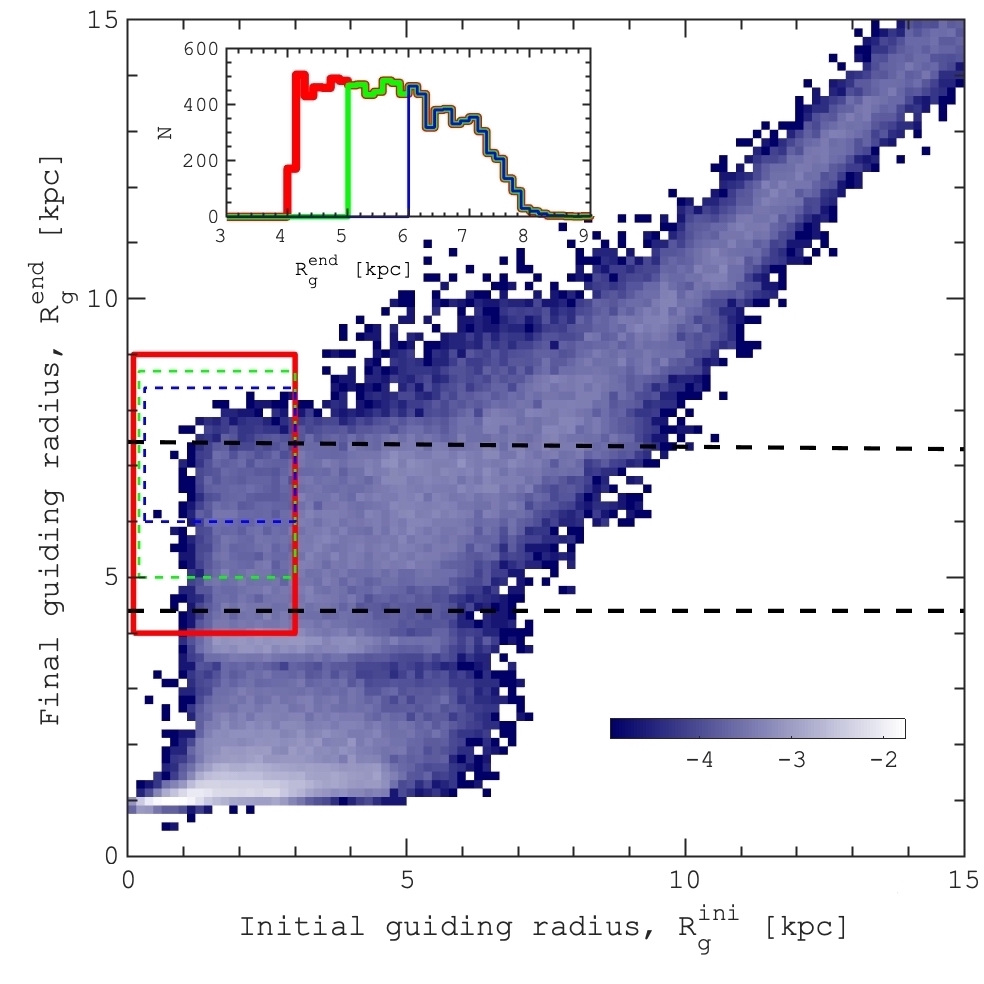}
\caption{Relation between the initial value  ($\Rgini$, the mean value at $T<250$~Myr) and the  final value ($\Rgend$, mean value at $6<T<7$~Gyr) of guiding radius for thin-disk stars. The color-coding indicates the fraction of stars in log-scale. Horizontal overdensity feature at $\Rgend \approx 4$~kpc is associated with the corotation radius of the bar at the end of the simulation. Horizontal dashed lines show the mean location of the OLR and CR resonances of the bar~(from top to bottom, respectively) averaged over the last Gyr of evolution. The  red box contains stars that migrate mostly from the inner region ($\Rgini<3$~kpc),  while the green and blue boxes correspond to the selection weighted toward larger final guiding radii. The inset shows the distribution of guiding radii of all the stars in the red, green, and blue boxes at the end of simulation~($\Rgend$). The statistical properties~(mass, distribution of orbital parameters) of the inner disk migrators are  dependent on the exact selection of a red region in the diagram.}\label{fig::rg_ini_rg_fin}
\end{center}
\end{figure}

\section{Model}\label{sec::models}
We explore a purely collisionless $N$-body simulation of a disk galaxy with a total stellar mass and a rotation curve compatible with those of the Milky Way~(see Fig.~\ref{fig::app1}). In this work we employ a multicomponent model for a galaxy consisting of three co-spatial disk populations: cold, warm, and hot disks represented by $5 \times 10^6$, $3 \times 10^6$, and $2 \times 10^6$ particles respectively. Initially stellar particles redistributed following a Miyamoto--Nagai density profile~\citep{1975PASJ...27..533M} that has a characteristic scale length of $4.8$, $2$, and $2$ kpc; vertical thicknesses of $0.15$, $0.3,$ and $0.6$~kpc; and masses of $4.21$, $2.57,$ and $1.86 \times 10^{10}$~\Msun, respectively. The kinematically warm and hot disk components have a mass comparable to that of the cold thin disk, in agreement with the finding that the chemically defined thick disk of the Galaxy is massive, as shown by~\citet{2014ApJ...781L..31S} where the Milky Way thick disk represents about a half of the disk mass within $10$~kpc of the galactic center.  Our simulation also includes a live dark matter halo~($5\times 10^6$ particles) whose density distribution follows a Plummer sphere~\citep{1911MNRAS..71..460P}, with a total mass of $3.81\times 10^{11}$~\Msun and a radius of $21$~kpc. The choice of parameters leads to a galaxy mass model with a circular velocity of $\approx220$~\kmps and Toomre stability parameter $Q_T\approx1.5$ at $8$~kpc. The initial setup was generated using the iterative method by~\citet{2009MNRAS.392..904R}.

This model has been already successfully used to  reproduce the morphology of the metal-rich and metal-poor stellar populations in the Milky Way bulge, as well as the bulge mean metallicity and $\rm [\alpha/Fe]$ maps  obtained from the APOGEE data~\citep{2017A&A...607L...4F,2018A&A...616A.180F}. It has also been employed to study and interpret the complex phase-space structures recently discovered in the Milky Way~\citep{2019A&A...622L...6K,2019MNRAS.488.3324F}.  

\begin{figure}[t!]
\begin{center}
\includegraphics[width=1\hsize]{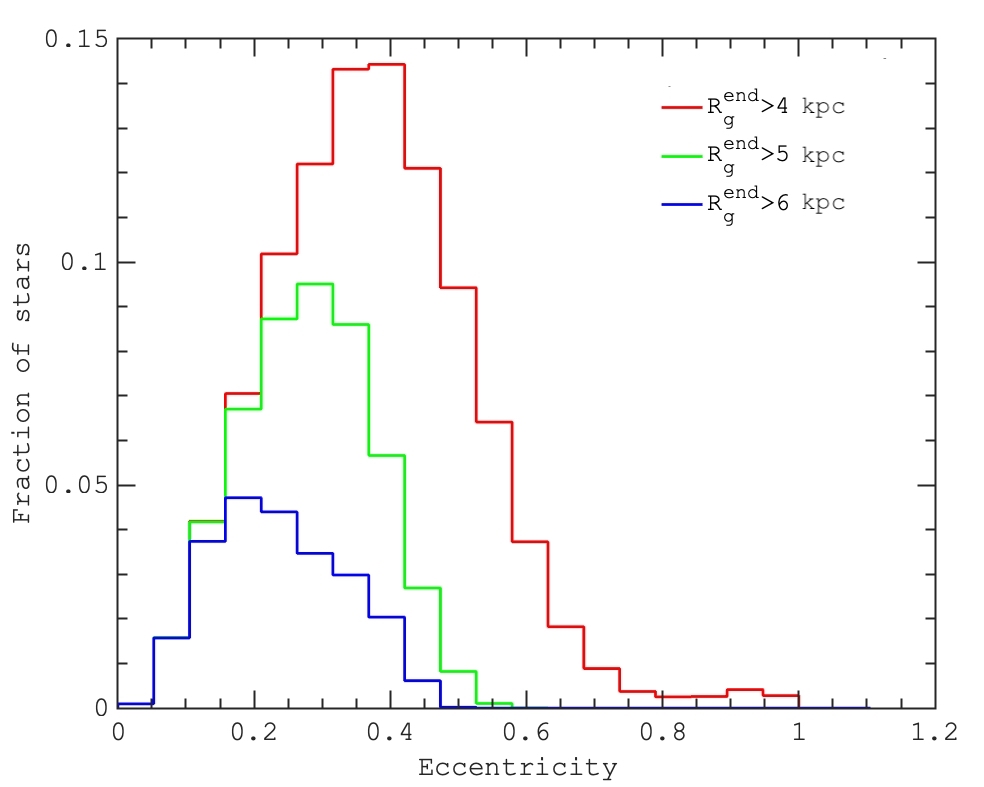}
\caption{ Distribution of orbital eccentricities~(see Eq.~\ref{eq::ecc}) of migrating stars selected in red, green, and blue boxes in Fig.~\ref{fig::rg_ini_rg_fin}. The eccentricities are calculated by using apocenters and pericenters of stellar orbits over the last $1$~Gyr of the simulation.}\label{fig::diff_rgend_ecc}
\end{center}
\end{figure}

In this simulation, the bar starts to form in the first Gyr of evolution, and by $\approx1$~Gyr, it reaches its maximum strength. Its strength remains nearly constant for about $4$~Gyr (from $1$ to $5$~Gyr, see Fig.~\ref{fig::app1}, top right panel). At $t = 5$~Gyr, the bar strength starts to decline because of the vertical buckling of the bar and the subsequent formation of a boxy/peanut bulge, as  has been reported in a number $N$-body simulations~\citep[][]{1981A&A....96..164C, 1990A&A...233...82C, 2006ApJ...637..214M, 2008MNRAS.390L..69A,2019A&A...628A..11D,2020arXiv200206627P,2020arXiv200300015C}. As a result of the bar slowdown~\citep[see][]{2000ApJ...543..704D, 2003MNRAS.341.1179A}, its length increases and thereby its resonances migrate outward~\citep[see, e.g.,][]{2018A&A...616A..86H}. The temporal evolution of the bar strength ($A_2$), of its pattern speed, and of the location of bar resonances is shown in Fig.~\ref{fig::app1}. In this figure, the location of the resonances is calculated by making a Fourier analysis of the face-on stellar density maps at different times, and therefore the location of the bar resonances in Fig.~\ref{fig::app1} must be seen as a  mean location because the whole resonance region can be very wide in radius, as shown when an analysis of individual orbits of stars is performed~\citep{2007MNRAS.379.1155C,2018A&A...616A..86H,2019arXiv191006335K}.

\section{Results}\label{sec::results}

In this section, we analyze the evolution of our $N$-body model, and, in particular, the change in orbital parameters of stars due to the bar slowdown. In particular, we focus on the analysis on the orbital evolution of stars placed, before the formation of the bar, in the inner parts of the disk, which -- as we recall  in Sect.~\ref{sec::obs} -- should constitute the original birthplace of the high-metallicity stars currently found  at the solar vicinity and beyond. 

\begin{figure}[t!]
\begin{center}
\includegraphics[width=1\hsize]{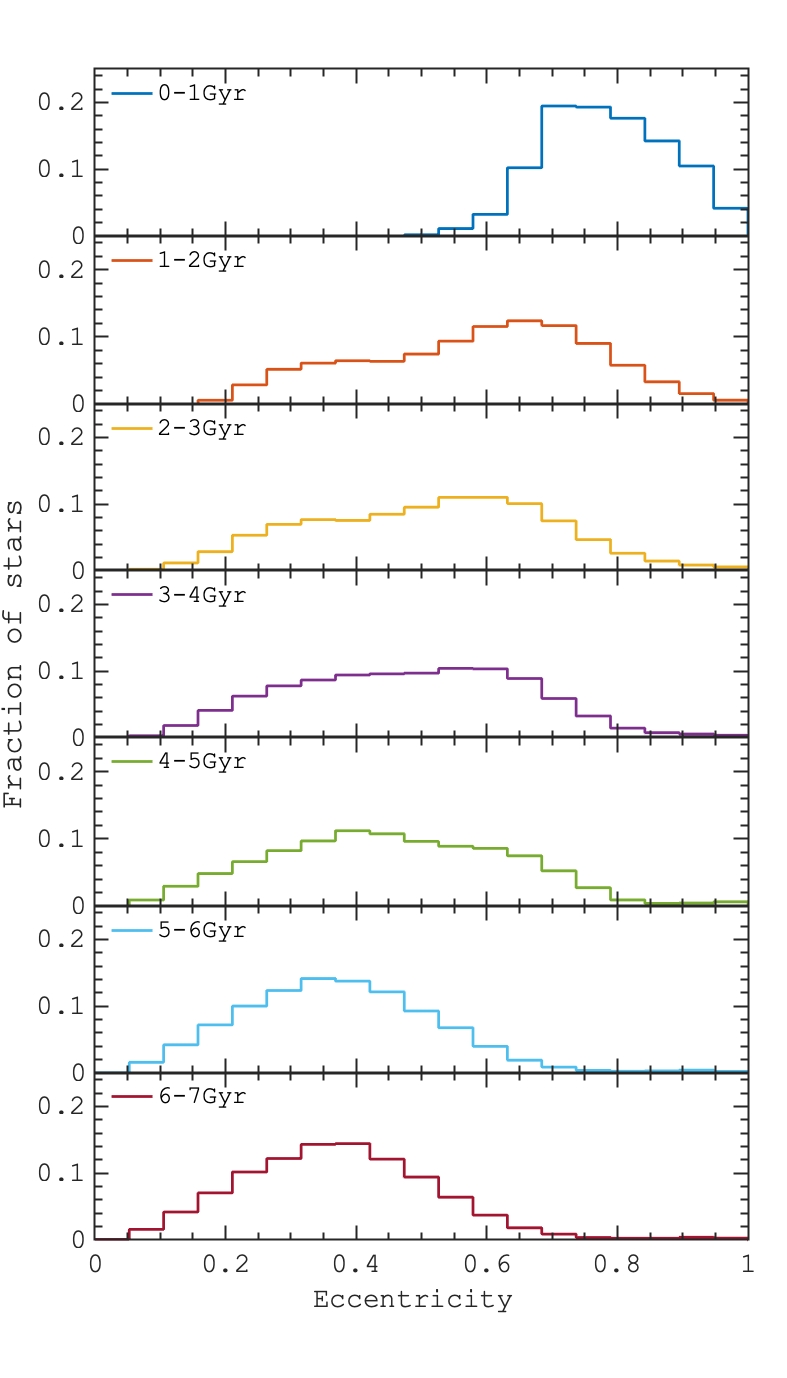}
\caption{ Distribution of orbital eccentricities~(see Eq.~\ref{eq::ecc}) of migrating stars selected in the red box in Fig.~\ref{fig::rg_ini_rg_fin}. The eccentricities are calculated by using apocenters and pericenters of stellar orbits measured in different $1$~Gyr time intervals, as   mentioned in each frame. The mean value of eccentricity tends to decrease in time, and low-eccentricity stars start to appear after $3$~Gyr. }\label{fig::distr_ecc}
\end{center}
\end{figure}

\subsection{Migration in the barred galaxy: global picture}

Figure~\ref{fig::rg_ini_rg_fin} shows the relation between the initial guiding radii, \Rgini, of stars in the simulated disk, and their final guiding radii, \Rgend. As in \citet{2015A&A...578A..58H}, guiding radii~($R_g$) are estimated as  the mean value between minima and maxima of the oscillatory evolution of the radius $R$ (i.e., of the in-plane distance from the galactic center) for all individual particles. We note  that in our simulations the separation between two output snapshots is $10$~Myr, which guarantees a   temporal resolution high enough to apply the method of the guiding radii calculations, described in detail in \citet{2015A&A...578A..58H}. The initial guiding radii are estimated by averaging $R_g$ over the first $250$~Myr of the disk evolution (when the disk is still axisymmetric and the bar is not formed yet); the final guiding radii are estimated by averaging $R_g$ over the last $1$~Gyr of the simulation, that is in the time interval $6<t<7$~Gyr. Because a change in  guiding radius corresponds to a change in angular momentum,  the relation between these two estimates can be used to quantify radial migration~(churning). 

Figure~\ref{fig::rg_ini_rg_fin}  clearly demonstrates that stars in the inner galaxy can migrate in both  inward~(i.e., $ \Rgend-\Rgini < 0$) and outward~(i.e., $ \Rgend-\Rgini > 0$). At the epoch of the bar formation~($t<1$~Gyr), and during the early phase of strong bar activity~($1<t<2$~Gyr), stars in the disk experience a significant spatial redistribution, with a probability of migration maximum at the bar corotation resonance which is demonstrated in our $N$-body model and in agreement with some previous studies~\citep{2011A&A...527A.147M, 2013A&A...553A.102D}. In particular, in Figure~\ref{fig::rg_ini_rg_fin}, the $\Rgini-\Rgend$ plane shows a significant horizontal overdensity at $R_g \approx 4$~kpc, just inside  the locus of the bar corotation at the end of the simulation. The most intense outward migration is experienced by stars whose initial guiding radius is at $\Rgini \approx 2$~kpc, which coincides with the location of the bar corotation at early times (see below). These stars can migrate as much as about $5$~kpc, reaching a final guiding radius $\Rgend \approx 7$~kpc.

 Figure~\ref{fig::rg_ini_rg_fin} also shows that the distribution in the $\Rgini-\Rgend$ plane is truncated at $\approx 7.4$~kpc~(see inset), which corresponds to the mean location of the OLR at the end of the simulation.  In greater detail, the absence of stars with $\Rgini < 4$~kpc and $\Rgend > 8$~kpc reflects the fact that stars whose guiding radii are, at early times, inside the bar OLR~(initially located at $R_g \approx 4$~kpc, see Fig.~\ref{fig::rg_time_map}) cannot propagate their guiding radii outside the bar OLR at any following time; in other words, the bar OLR acts as a natural barrier keeping stars that are  initially inside it, always inside it~\citep[see also][]{2015A&A...578A..58H}. In this respect, it should not be surprising to find a tiny fraction of stars beyond $7.4$~kpc because the bar OLR (and the corotation) region can be broad in galactocentric radii, and our estimates (e.g., in Fig.~\ref{fig::app1}) provide only its mean location, averaged over the last Gyr of evolution.

\begin{figure}[t!]
\begin{center}
\includegraphics[width=1\hsize]{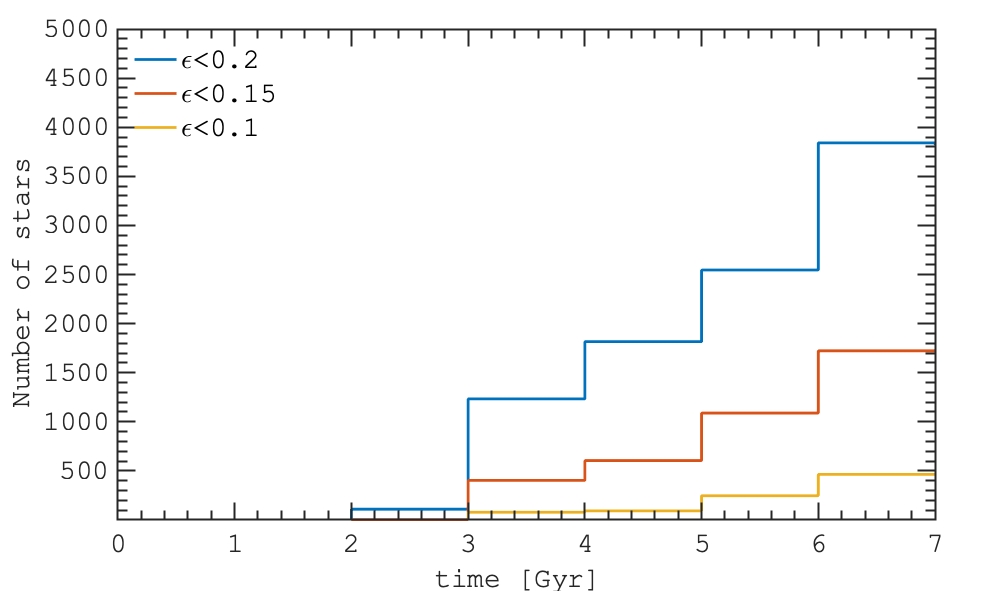}
\caption{ Number of stars with low orbital eccentricities as a function of time for stars selected in the red box in Fig.~\ref{fig::rg_ini_rg_fin}. The eccentricities are calculated by using apocenters and pericenters of stellar orbits  measured in different $1$~Gyr time intervals.}\label{fig::ecc_time}
\end{center}
\end{figure}

\begin{figure*}[t!]
\begin{center}
\includegraphics[width=1\hsize]{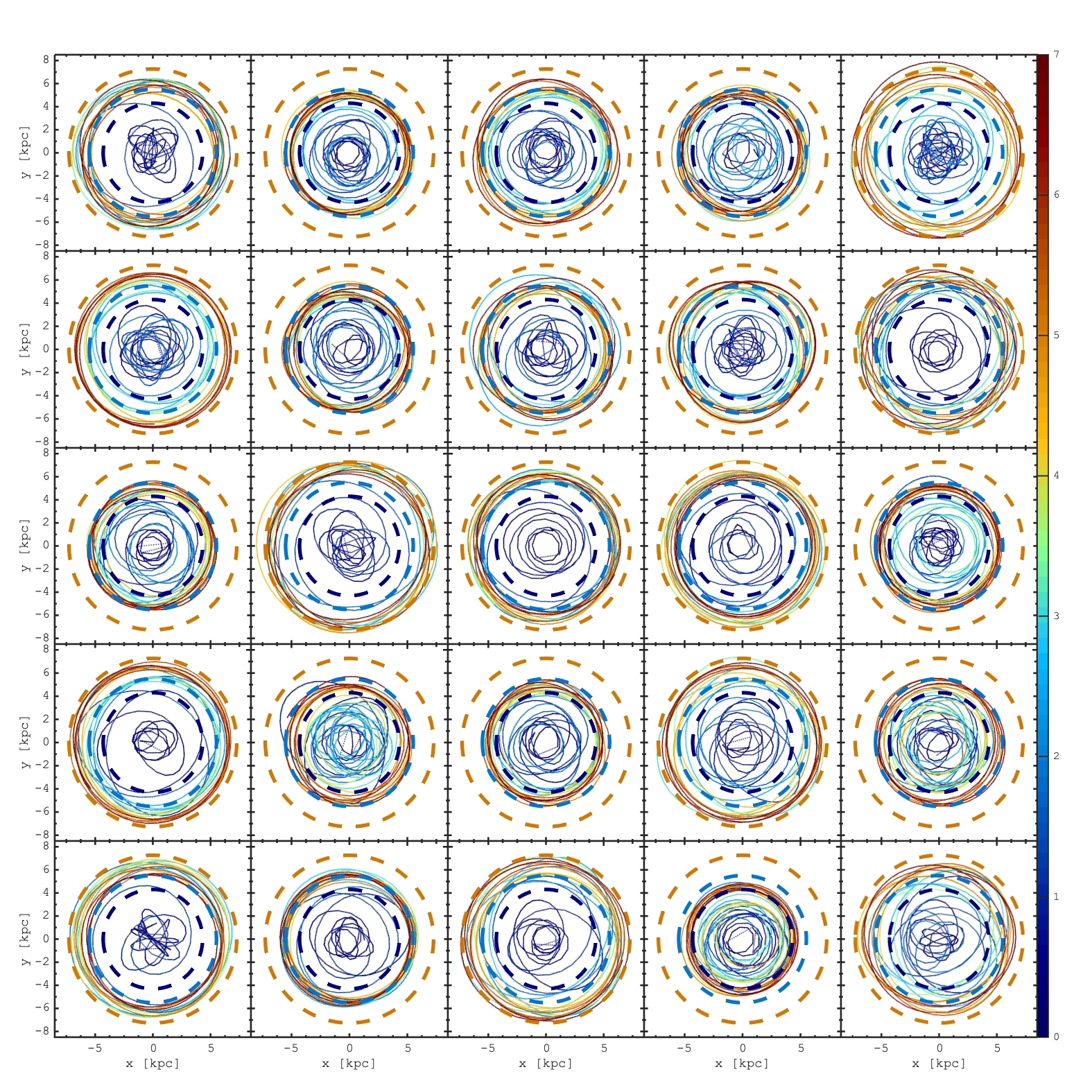}
\caption{Orbits of stars migrated from the inner galaxy and have nearly circular orbits at the end of the simulation. Orbits are  color-coded by time in Gyr; the dashed circles from inner to outer show a position of the OLR at $0.5$, $2,$ and $7$~Gyr. The color bar shows the time for  orbits of stars and for the OLR in units of Gyr. Each frame box size is 16~kpc $\times$ 16~kpc.}\label{fig::app2}
\end{center}
\end{figure*}

\subsection{Finding the inner disk migrators}\label{sec::sec42}

In this work our aim is to test the possibility that high-metallicity stars currently found  at the solar vicinity were formed in the inner disk, and migrated outward following the resonances of a slowing-down bar. Thus, as a first step we need to select a region of the simulated disk to compare with observations. This is not trivial since the exact location of the Sun with respect to the Milky Way bar resonances is still under debate: a number of works have suggested that the solar radius is just beyond the bar OLR~\citep{2000AJ....119..800D,2007ApJ...664L..31M,2003ApJ...599..275O,2010MNRAS.407.2122M,2019MNRAS.488.3324F}, while some others claim that the bar OLR is outside the solar radius, and that  as a consequence  the Sun should lie between the bar CR and OLR~\citep{2013MNRAS.435.1874W, 2017MNRAS.465.1621P,2019MNRAS.488.4552S,2019MNRAS.489.3519C,2020A&A...634L...8K}. 

In our analysis we can thus operate a very broad selection by analyzing all stars whose guiding radii, at the end of the simulation, are located between the final CR and inside the final OLR, but which have, at the beginning of the simulation, guiding radii confined in the innermost disk (see red box in Figure~\ref{fig::rg_ini_rg_fin}). Among these stars,  in particular we aim to identify and study those that, at the end of the simulation, can be found on quasi-circular orbits. Observations indeed show that some of the metal-rich stars found at the solar vicinity  have very low eccentricities, even though they are too metal rich to have formed at the solar radius~(see Sect.~\ref{sec::obs}). If we are able to show that stars trapped at the resonances (and in particular at the OLR) of a slowing-down bar can have orbital properties consistent with those of metal-rich stars in the observations, this will imply that the mechanism under study in this paper is able to explain the existence of these stars at the solar vicinity.

 It should be noted that the choice of the exact values of \Rgend and \Rgini relative to the bar resonance locations that we adopt in this paper is somewhat arbitrary, and any comparison with observations should take into account this aspect carefully.
To demonstrate this, we first consider three disk regions containing migrated stars: the red box ($\Rgend>4$), green box ($\Rgend>5$), and blue box~($\Rgend>6$)~(see Fig.~\ref{fig::rg_ini_rg_fin}). 

In order to study the orbital structure of these migrators, we use the eccentricity parameter based on the calculation of the pericenter~($\rm R_{min}$) and apocenter~($\rm R_{max}$) for individual star particles:
\begin{equation}
\rm \varepsilon = 1 - \frac{2}{R_{max}/R_{min}+1},\label{eq::ecc}
\end{equation}
 where $R_{min}$ and $R_{max}$ are respectively the pericenter and apocenter of its orbit, which are calculated as the minimum and maximum values of the galactocentric radius~($R$) during a certain time interval. In Figure~\ref{fig::diff_rgend_ecc} we present the eccentricity distribution for migrating stars that are located   between the corotation and the OLR for different choices of the minimum value of the final guiding radius. The mean eccentricity of migrating stars decreases as we restrict our analysis to regions far away from the corotation, and reaches about $0.2$ for stars with final guiding radii between 6 and 8.5~kpc. This occurs because when the chosen region is farther from corotation, it contains fewer stars trapped at corotation on highly elongated orbits. There is thus a dependence of the results presented in the following on the exact choice of the region under analysis. To describe the wider possible range of stellar parameters, we focus our analysis on the greater guiding radii range~(red box).

\subsection{Temporal evolution and orbits of the inner disk migrators}

In Fig.~\ref{fig::distr_ecc} we present the distribution of eccentricities for all the particles located in the selected region of the  $\Rgini-\Rgend$ plane (that is $\Rgini < 3$~kpc, and $4 < \Rgend < 9$~kpc; see red box in Fig.~\ref{fig::rg_ini_rg_fin}). In the early phases of the galaxy evolution, during the epoch of the bar formation and the short phase of strong bar activity~($t<2$~Gyr), the orbits of the selected stars are characterized by large radial excursions~(eccentricities $\langle \varepsilon \rangle \approx 0.8$) because of the rapid change in their guiding radii from small values of $R_g$, at the beginning, to larger $R_g$ at later times. During the epoch of the steady bar evolution, characterized by a nearly constant bar strength~($1<t<5$~Gyr), the distribution of eccentricities slowly shifts to lower mean values. It is interesting to note that already during the steady epoch of the bar evolution, and later on after its vertical buckling~($t>5$~Gyr), the number of stars on quasi-circular orbits gradually increases. The presence of low-eccentricity stars at the solar radius is not obvious a priori because stars in the inner galaxy tend to have higher velocity dispersions and, once moved outward to regions with lower disk gravity, they should keep showing very large radial~(and vertical) excursions with relatively high orbital eccentricities~\citep[see, e.g.,][]{2014A&A...572A..92M, 2015ApJ...804L...9M}. 

\begin{figure}[t!]
\begin{center}
\includegraphics[width=1\hsize]{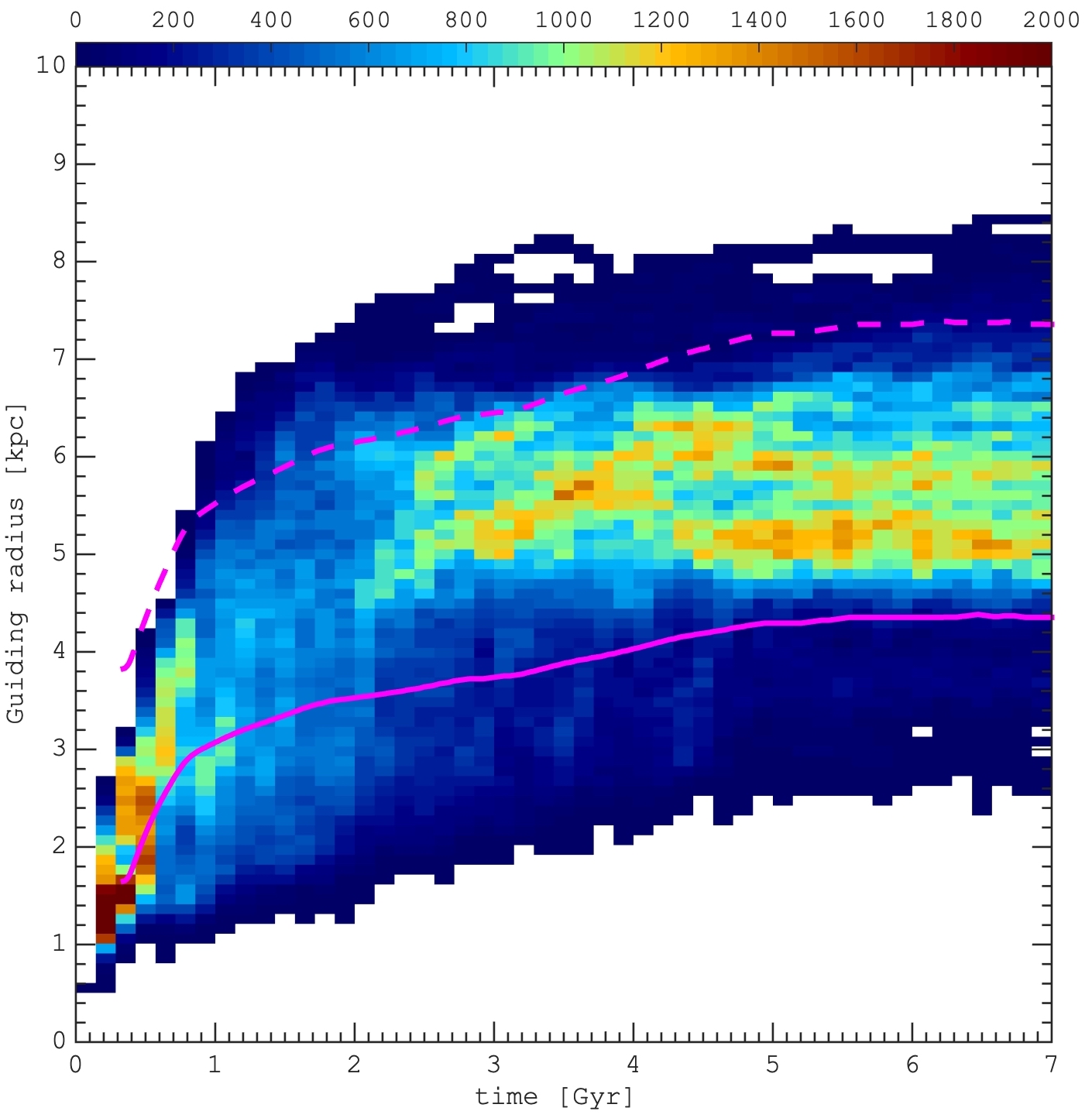}
\caption{ Superposition of orbits~(galactocentric distance $R$ as a function of time) of stars that migrate from the inner region~($\Rgini<3$~kpc) and have a low orbital eccentricity~($\varepsilon<0.15$) at $t=7$~Gyr. The number of stars in a given time-radius bin is indicated by color (see   scale at top). The intensity of migration is strongly connected to the slowdown of the bar and to the following propagation of its resonances outward. Most of the migrating stars are locked   between two resonances: CR~(solid line) and OLR~(dashed line).}\label{fig::rg_time_map}
\end{center}
\end{figure}

The temporal evolution of the number of migrated stellar particles on low eccentricities or circular orbits~($\varepsilon<0.1; 0.15; 0.2$) is shown in Fig.~\ref{fig::ecc_time}. These low-eccentricity migrators are redistributed over a large portion of the simulated disk~(wide range of guiding radii). Although the observed gradual decrease in the orbital eccentricity can be caused by an increase in the guiding centers of the orbits~(see Eq.~\ref{eq::ecc}), next we illustrate that the bar-induced outward migration leads to truly nearly circular orbits with small radial motions. In particular, the orbits of some of these migrators,  randomly selected among those with final eccentricity  $\varepsilon<0.15$,   are presented in Fig.~\ref{fig::app2}. These orbits are color-coded by time from blue~(at the beginning of simulation, $t=0$) to brown~(at the end of the simulation, $t=7$~Gyr). This figure clearly shows the orbit circularization of the orbits of some of the migrators: initially elongated in the inner disk, and increasingly circular as they move outward.

During the long secular evolution phase~($2<t<5$~Gyr) most of the low-$\varepsilon$ migrating star particles redistribute midway between the bar CR and OLR. More detailed changes in the radial distribution of the stars can be found in Fig.~\ref{fig::rg_time_map}, where we present the distribution of guiding radii as a function of time for stellar particles that have a low eccentricity ($\varepsilon<0.15$) at the end of the simulation. We note that the migration is stronger at the epoch of bar growth and it is less significant at later times. Therefore, {inner disk stars can be deposited~(churned) to large guiding radii due to the gain in angular momentum at the epoch of formation of the bar. In this respect, the population of very metal-rich stars found at the solar vicinity can represent the fossil signature of the bar-induced migration occurred at the epoch of its formation.

\subsection{Origin of low-eccentricity migrators in the bar slowdown model}

To further probe that the stellar particles which at the end of the simulation are on quasi-circular orbits, and which were initially in the inner disk, have migrated because they are trapped at resonances, we carry out a spectral analysis of the orbits~\citep[see also][]{1982ApJ...252..308B, 2007MNRAS.379.1155C, 2018A&A...616A..86H} in order to obtain the azimuthal and radial orbital frequencies, $\Omega$ and $\kappa$, for these stars. To further probe the stellar particles (which were initially in the inner disk and at the end of the simulation were on quasi-circular orbits) that have migrated because they are trapped at resonances, we carried out a spectral analysis of the orbits~\citep[see also][]{1982ApJ...252..308B, 2007MNRAS.379.1155C, 2018A&A...616A..86H} in order to obtain the azimuthal and radial orbital frequencies, $\Omega$ and $\kappa$, for these stars.

Figure~\ref{fig::freq} shows the distributions of the ratio $(\Omega-\Omega_{b})/\kappa$, where $\Omega_{b}$ is the bar pattern speed measured by averaging its value over time intervals of  $1$~Gyr  amplitude. For each time interval, as shown in the figure, we compare the distribution of stars migrated from the inner region~(\Rgini<3) on circular orbits~($\varepsilon<0.5$, colored lines) at the end of the simulation with the distribution of $200000$  randomly selected stellar particles, from the disk~(black lines). The most prominent peaks correspond to the main resonances of the bar~($(\Omega-\Omega_{b})/\kappa=0$ at corotation and $(\Omega-\Omega_{b})/\kappa = \pm 0.5$ at the inner and outer Lindblad resonance, respectively). 

\begin{figure*}[t!]
\begin{center}
\includegraphics[width=1\hsize]{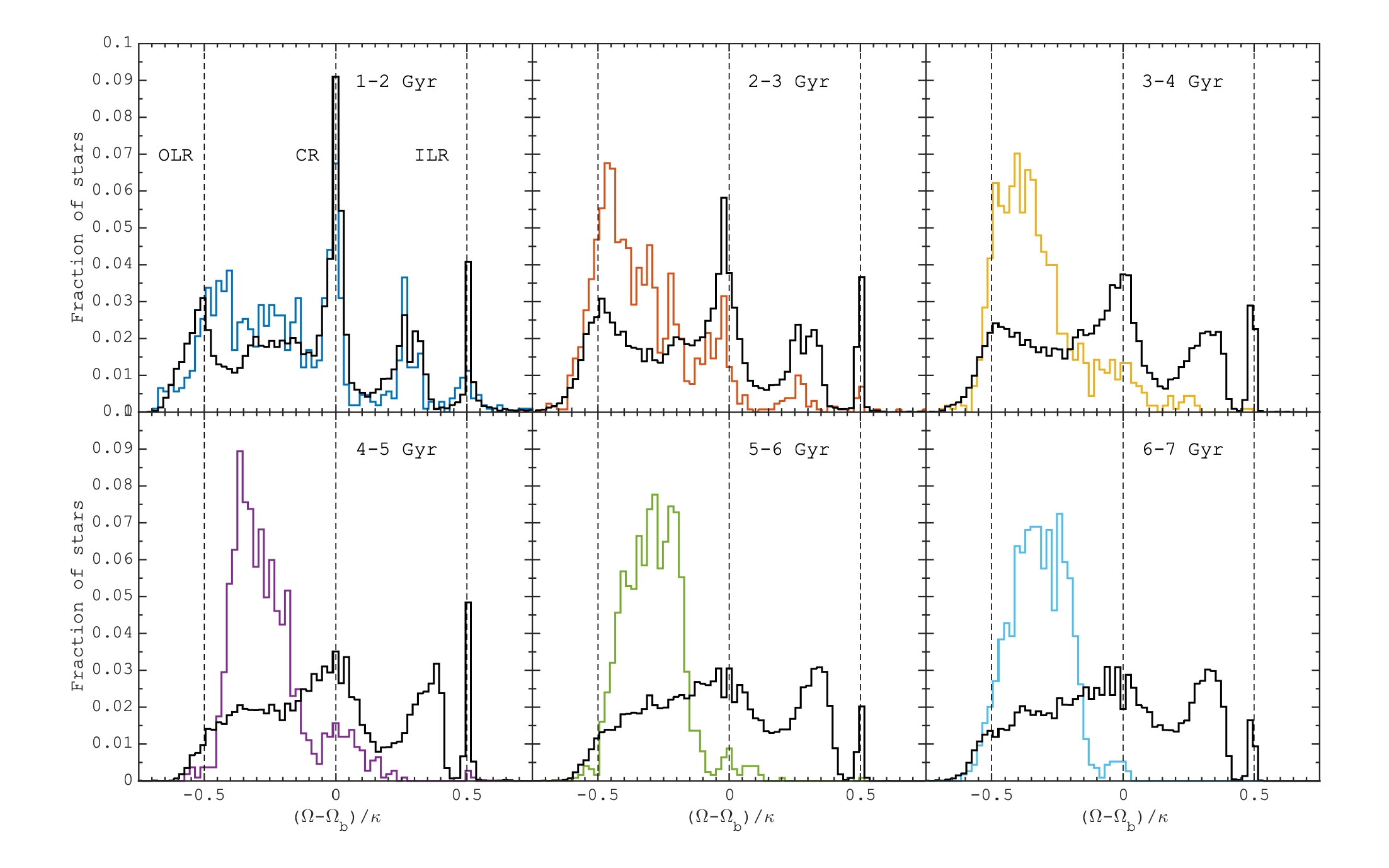}
\caption{Frequency ratio of orbits of migrated stars with $\varepsilon<0.5$ at $7$~Gyr shown by the colored lines for different periods. For the same time intervals the  frequency ratios for 200000 randomly selected particles in the whole disk are shown by the black lines. Both distributions were normalized to the total number of stars in each sample separately.  The vertical dashed lines indicate three main resonances of the bar: the ILR, the CR, and the OLR (from left to right). Diminishing of the bar strength over time is depicted in the decrease in a number of stars associated with its main resonances. Migrated low-$\varepsilon$ stars almost escaped the resonances at the end of the simulation.}\label{fig::freq}
\end{center}
\end{figure*}

At the epoch of bar formation and during the short phase of the bar strengthening~($t<2$~Gyr), low-eccentricity particles (particles that have low eccentricity at the end of simulation) from the inner disk show the same relation to the resonances as the rest of the disk. At $t>2$~Gyr these particles tend to be trapped at the bar OLR and corotation, while the disk on average does not change its resonant structure drastically. Once the bar strength diminishes ($t>5$~Gyr), the migrated stars tend to escape the main bar resonances and occupy the region  between corotation and the OLR. In particular, OLR-stellar particles escape entirely from the resonance at $4-5$~Gyr, while those at corotation also leave the resonance, but on a longer timescale. This process demonstrates how migrators from the inner disk that end up on circular orbits populate the region  between  the corotation and the OLR~(see Fig.~\ref{fig::rg_time_map}). 

Throughout this work, we restrict our analysis to migration from the innermost disk because we aim to explore the origin of high-metallicity stars in the solar radius. These stars are the most extreme migrators since they can be moved up to $\sim 6-7$~kpc away from their birthplaces~(see Fig.~\ref{fig::app2}). Meanwhile, a significant fraction of stars formed within the OLR can also experience a moderate migration~(a few kpc)  due to the bar slowdown mechanism. In Table~\ref{table}, we summarize the main characteristics of migrated inner disk stellar populations to be compared to those observed in the Milky Way. We find that due to the bar slowdown mechanism about $1.4-3.3$\% of the thin disk mass   between the bar corotation and the OLR is represented by stars formed in the inner galaxy  which is in agreement with available estimates of the fraction of  metal-rich stars  in the solar vicinity~\citep{2018A&A...609A..79H,2019A&A...625A.105H}. Among these migrators, the fraction of stars on nearly circular orbits~($\varepsilon<0.2$) is between $10\%$ and $30\%$, depending on the choice of the region (see Table~\ref{table} and Fig.~\ref{fig::rg_ini_rg_fin}). This fraction also depends   on the time since the bar formation, and on its strength. As emphasized at the end of Sect.~\ref{sec::sec42}, these numbers should be taken with care, since they depend on the chosen region. For example, if instead of looking for all migrators with initial    guiding radius $\Rgini$ inside 3~kpc we had looked for all migrators with the same final guiding radii as defined by the red box in Fig.~\ref{fig::rg_ini_rg_fin} but initial guiding radii smaller than $4-5$~kpc, the fraction of migrators from the inner disk, relative to the thin disk mass  between the bar resonances, would have been substantially larger.

\begin{table}
\caption{Table summarizing the number of particles, masses, and fraction of the inner disk migrators ($\Rgini<3$~kpc) in particular disk region~($\Rgend$ range) as   indicated in Fig.~\ref{fig::rg_ini_rg_fin}: red~($4<\Rgend<8$~kpc), green~( $5<\Rgend<8$~kpc), and blue~( $6<\Rgend<8$~kpc) boxes.}
\begin{tabular}{lccccccccccccccccccc}
\hline          
 & Number & Mass & Fraction  \\
& of particles &  \Msun &  \% \\
\hline          
\hline          
Whole disk & $10\times10^6$ &  $8.64 \times 10^{10} $ & - \\
Thin disk & $5\times10^6$ &  $4.21 \times 10^{10} $ & - \\
\hline          
\hline
Red box selection & $4<\Rgend<8$ \\
\hline
All stars  & 1036112 & $8.7 \times 10^{9}$ &  -  \\
Inner disk migrators & 33851 &  $2.85\times10^8$ & 3.3 \\
$\varepsilon<0.2$ & 2761 &  $0.28\times 10^{8} $ & 0.3  \\
$\varepsilon<0.15$ & 1366 & $1.4\times 10^{7} $ & 0.13  \\
$\varepsilon<0.1$ & 428 &  $4\times 10^{6} $ & 0.04 \\
\hline
Green box selection &  $5<\Rgend<8$ \\
\hline
All stars  & 800473 & $6.7 \times 10^{9}$ & - \\
Iinner disk migrators & 11080 &  $9.3\times10^7$ & 1.4 \\
 $\varepsilon<0.2$ & 3515 &  $2.9\times 10^{7} $  & 0.44 \\
$\varepsilon<0.15$ & 896 & $7.5\times 10^{6} $  & 0.1 \\
 $\varepsilon<0.1$ & 316 &  $2.6\times 10^{6} $  & 0.04 \\
\hline
Blue box selection & $6<\Rgend<8$ \\
\hline
All stars  & 458420 & $3.8 \times 10^{9}$ & - \\
Inner disk migrators & 8583 &  $7.2\times10^7$ & 1.9 \\
$\varepsilon<0.2$ & 1973 &  $1.7\times 10^{7} $  & 0.43\\
$\varepsilon<0.15$ & 976 & $8.2\times 10^{6} $  & 0.2\\
$\varepsilon<0.1$ & 304 &  $2.5\times 10^{6} $  & 0.07\\
\hline          
\end{tabular}\label{table}
\end{table}

\section{Conclusions}\label{sec::discussion}
Radial migration was initially   proposed as a mechanism able to displace stars via angular momentum exchange near the spiral arms corotation resonance~\citep{2002MNRAS.336..785S}, displacement which occurs without changing the orbital eccentricity~\citep{2008ApJ...684L..79R,2012MNRAS.426.2089R}. Therefore, this particular mechanism of migration is more effective for stars on nearly circular orbits. 

In this work, by means of dissipationless $N$-body simulation, we studied the radial migration induced by the evolution of a stellar bar. Building on earlier works by~\citet{2015A&A...578A..58H, 2018A&A...616A..86H}, we studied the migration induced by a slowing down bar, on stars trapped at its main resonances. We showed that this mechanism is able to move stars from eccentric orbits in the inner disk to nearly circular orbits   reaching the outer disk regions. This mechanism is able to explain the presence of metal-rich stars ($\FeH > 0.2-0.3$) found at the solar vicinity, in a framework where these stars migrated from the inner disk at the time of the formation of the bar, having being trapped at its resonances during its growth and slowdown. Because the distribution of stars at resonances is rather wide~\citep{2018A&A...616A..86H,2019arXiv191006335K}, we expect that in the Milky Way stars that experienced radial migration associated with the bar resonances migration should be found in a wide range of galactocentric radii.

Another possible implication of our results is related to the possible estimation of the bar formation epoch in the Milky Way. We demonstrated that the most intense migration appears at the epoch of the bar formation, which is followed by a rapid decrease in its pattern speed and migration of the bar resonances. We note that signatures of the bar deceleration have been potentially found in the Milky Way, by studying local stellar kinematics~\citep{2019arXiv191204304C}, thus making our proposed scenario particularly suitable for the Milky Way. Therefore, we suggest that the age of the youngest high-metallicity stars~($\FeH>0.2-0.3$) on nearly circular orbits in the solar vicinity should correspond to the age of the Milky Way bar. These stars are likely to be the latest populations formed in the galactic center before the bar formation, and together with even older populations they can efficiently be transferred outward. Later populations formed in the galactic center will not be churned out by the bar if it  experiences steady evolution, as  is expected from $N$-body simulations~\citep[see Fig.~\ref{fig::app1} and, e.g.,][]{2000ApJ...543..704D,2003MNRAS.341.1179A,2009ApJ...697..293D}. 

\begin{acknowledgements}
The authors are grateful to the referee, for their very constructive report which improved the manuscript. Authors also thank  Victor Debattista for several useful comments. This work was granted access to the HPC resources of CINES under the allocation 2017-040507 (PI : P. Di Matteo) made by GENCI. PDM and MH thank the ANR (Agence Nationale de la Recherche) for its financial support through the MOD4Gaia project (ANR-15- CE31-0007, P.I.: P. Di Matteo). Numerical simulations were carried by using the equipment of the shared research facilities of HPC computing resources at Lomonosov Moscow State University supported by the project RFMEFI62117X0011. 
SAK acknowledges support from Russian Science Foundation~(project no. 19-72-20089). ONS acknowledges DIM ACAV+ funding. 
\end{acknowledgements}

\bibliographystyle{aa}
\bibliography{references-1}

\end{document}